\documentclass[preprint]{aastex}
\usepackage{emulateapj5}
\usepackage{apjfonts}
\usepackage{natbib}
\input psfig.tex

\def\aa{{A\&A}}

\def\aj{{AJ}}

\def\apj{{ApJ}}
\def\apjs{{ApJS}}

\def\etal{{et al. }}

\def\gax{{$\mathrel{\hbox{\rlap{\hbox{\lower4pt\hbox{$\sim$}}}\hbox{$>$}}}$}}

\def\hst{{\it HST}}
\def\kms{km s$^{-1}$}

\def\lax{{$\mathrel{\hbox{\rlap{\hbox{\lower4pt\hbox{$\sim$}}}\hbox{$<$}}}$}}
\def\lum{erg s$^{-1}$}

\def\mnras{{MNRAS}}

\def\pasp{{PASP}}

\def\percm2{cm$^{-2}$}

\def\solmass{$M_\odot$}

\def\caii{\ion{Ca}{2}}

\received{}
\accepted{}
\slugcomment{To appear in {\it The Astrophysical Journal (Letters).}}
\lefthead{FILIPPENKO and HO}
\righthead{LOW-MASS BLACK HOLE IN NGC 4395}
 
\begin{document}
 
\title{A Low-Mass Central Black Hole in the Bulgeless Seyfert 1 
Galaxy NGC 4395}
 
\author{
Alexei V. Filippenko\altaffilmark{1} and Luis C. Ho\altaffilmark{2}
}
 
\altaffiltext{1}{Department of Astronomy, University of California, Berkeley,
CA 94720-3411; alex@astro.berkeley.edu.}
 
\altaffiltext{2}{The Observatories of the Carnegie Institution of Washington,
813 Santa Barbara St., Pasadena, CA 91101; lho@ociw.edu.}
 
\begin{abstract}

NGC 4395 is one of the least luminous and nearest known type~1 Seyfert
galaxies, and it also lacks a bulge. We present a {\it Hubble Space Telescope
(HST)} $I$-band image of its nuclear region, and Keck high-resolution ($\sim$ 8
\kms) echelle spectra containing the \caii\ near-infrared triplet.  In addition
to the unresolved point source, there is a nuclear star cluster of size $r
\approx 3.9$ pc; the upper limit on its velocity dispersion is only 30 \kms.
We thus derive an upper limit of $\sim 6.2 \times 10^6~M_\odot$ for the mass of
the compact nucleus.  Based on the amount of spatially resolved light in the
{\it HST} image, a sizable fraction of this is likely to reside in stars.
Hence, this estimate sets a stringent upper limit on the mass of the central
black hole.  We argue, from other lines of evidence, that the true mass of the
black hole is likely to be $\sim 10^4-10^5$ \solmass.  Although the black hole
is much less massive than those thought to exist in classical active galactic
nuclei, its accretion rate of $L_{\rm bol}/L_{\rm Edd} \approx 2 \times
10^{-2}$ to $2 \times 10^{-3}$ is consistent with the mass-luminosity relation
obeyed by classical AGNs. This may explain why NGC 4395 has a high-excitation
(Seyfert) emission-line spectrum; active galaxies having low-ionization spectra
seem to accrete at significantly lower rates.  NGC 4395, a pure disk galaxy,
demonstrates that supermassive black holes are not associated exclusively with
bulges.

\end{abstract}
 
 
\section{Introduction}

The nucleus of NGC 4395, an Sd~III--IV galaxy (Sandage \& Tammann 1981) at a
distance of only $\sim 4.2$ Mpc (D. Minitti et al., in preparation)\footnote{We
scale all distance-dependent quantities to this value, revised from the
often-quoted distance of 2.6 Mpc from Rowan-Robinson (1985).}, exhibits strong,
narrow emission lines with relative intensities similar to those of type~2
Seyfert galaxies.  Spectra having high signal-to-noise ratios, however, reveal
the presence of faint, broad components to the permitted hydrogen and helium
lines; NGC 4395 therefore has a Seyfert~1 nucleus, one of the nearest and least
luminous known (Filippenko \& Sargent 1989). With $M_B \approx -10.8$ mag, it
provides an exceptional opportunity to study the active galactic nucleus (AGN)
phenomenon on luminosity scales comparable to those of the brightest
stars. Moreover, unlike the case in other low-luminosity AGNs (e.g., M81;
Filippenko \& Sargent 1988; Ho, Filippenko, \& Sargent 1996), starlight
contamination of the AGN in NGC 4395 is almost negligible because the bulge of
this dwarf galaxy ($M_B \approx -17.5$ mag) is essentially absent.  The extreme
late type of the host galaxy is highly unusual for AGNs, which tend to be found
in bulge-dominated systems (e.g., Ho, Filippenko, \& Sargent 1997).  The
nucleus of NGC 4395 is detectable as a variable X-ray source (Lira et al. 1999;
Moran et al. 1999; Iwasawa et al. 2000; Shih, Iwasawa, \& Fabian 2003; Moran et
al. 2003) that remains unresolved at {\it Chandra}\ resolution (Ho et
al. 2001).  It further exhibits a highly compact radio core (Moran et al. 1999;
Ho \& Ulvestad 2001) with a brightness temperature in excess of $2 \times 10^6$
K (Wrobel, Fassnacht, \& Ho 2001).

Here we present a careful search for stellar absorption lines in NGC 4395.  The
measured stellar velocity dispersion provides an upper limit to the mass of the
compact nucleus, and hence on the mass of the central black hole that
presumably powers the AGN.  This measurement has special significance in view
of the fact that NGC 4395 is a {\it bulgeless}\ galaxy.  Recent studies point
to the prevalence of central, supermassive ($10^6-10^9$ \solmass) black holes
and their close connection with galaxy bulges (Magorrian et al. 1998; Gebhardt
et al. 2000; Ferrarese \& Merritt 2000).  By contrast, supermassive black holes
seem to have little relation to galaxy disks (Kormendy et al. 2003).  Our mass
limit on NGC 4395 provides another critical data point for the currently poorly
constrained low-mass end of the mass spectrum of central black holes.

\section{Observations}

High-resolution ($R \equiv \lambda/\Delta\lambda =$ 38,000) spectra of the
nucleus of NGC 4395 were obtained on 1994 April 14 UT using the HIRES echelle
spectrometer (Vogt 1992; Vogt et al. 1994) at the Nasmyth focus of the Keck~I
10-m telescope; details will be presented in a subsequent paper
(A.~V. Filippenko \& L.~C. Ho, in preparation). Our setting spanned the
wavelength range 6338--8774~\AA\ (16 orders), and the slit width was 1\farcs15,
yielding a spectral resolution of $\sim 8$ \kms\ full width at half-maximum
intensity (FWHM).  Exposure times were 30, 60, and 60 minutes.  To enable
identification and removal of telluric lines, we also observed the sdF star HD
84937 (Oke \& Gunn 1983), whose spectrum has relatively few features.  Data
reduction followed standard procedures similar to those of Ho \& Filippenko
(1995).  The sky-subtracted, wavelength-calibrated, one-dimensional spectra
correspond to an effective aperture of 1\farcs15$\times$2\farcs05.

\section{Stellar Velocity Dispersion}

Our primary aim for the spectra was to detect, or set limits on, the \caii\
near-infrared triplet ($\lambda\lambda$8498, 8542, 8662). Figure~1 shows the
two relevant orders of the Keck echelle data for NGC 4395; the signal-to-noise
ratio per resolution element in the continuum is $\sim 35$. Also shown are
spectra of the 

\vskip 0.3cm
 
\begin{figure*}[t]
\centerline{\psfig{file=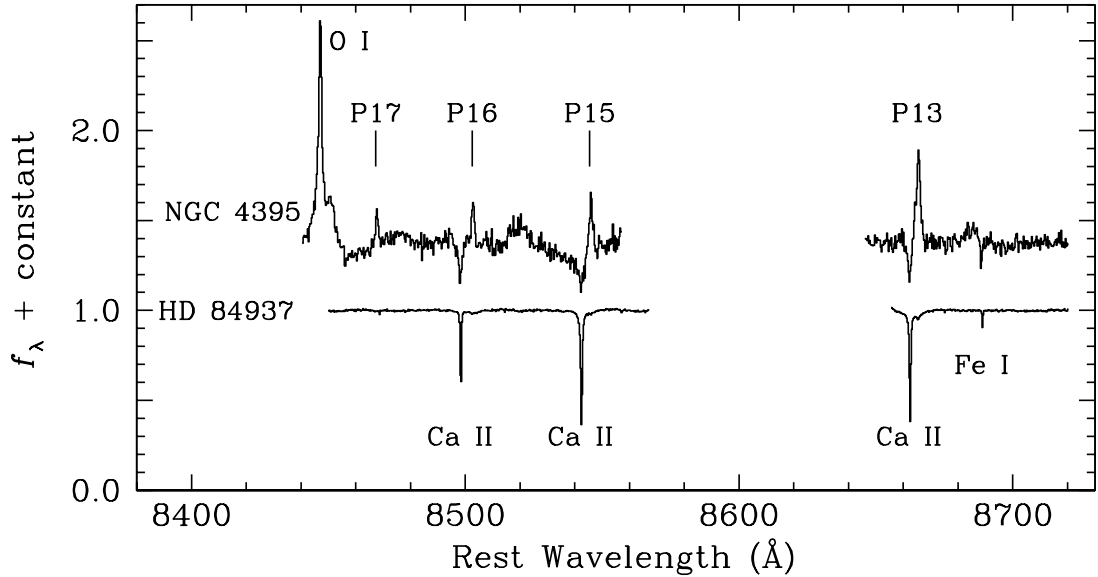,width=15.0cm,angle=270}}
\figcaption[fig1.ps]{
One order, and part of the next order, from the high-resolution spectrum
(FWHM = 8 \kms) of NGC 4395 obtained with the Keck~I 10-m telescope. Also
shown is the sdF comparison star HD 84937.  The gap between 8570 \AA\
and 8650 \AA\ is due to incomplete spectral coverage of the detector.
\label{fig1}}
\end{figure*}
\vskip 0.3cm

\noindent
comparison star HD 84937 taken with the same instrumental setup.
The \caii\ lines are detected in NGC 4395, at a systemic velocity of 320 \kms,
adjacent to the hydrogen Paschen emission lines.  These higher-order members of
the Paschen series are seldom seen in AGN spectra; they are detected in NGC
4395 because its recombination lines have exceptionally high equivalent widths.
Their presence, however, presents a challenge for measuring the \caii\ lines in
NGC 4395, since both components have comparable strengths.  The differences
between the laboratory wavelengths of the \caii\ lines and the adjacent Paschen
lines are 4.5, 3.3, and 2.8~\AA\ for \caii\ $\lambda\lambda$8498, 8542, and
8662 (respectively), and the FWHM of the Paschen lines is $\sim 1.7$~\AA.
Judging by the relative strengths and the profiles of the three \caii\ lines,
it appears that the $\lambda$8542 and $\lambda$8662 lines are substantially
affected by Paschen emission.  By simultaneously fitting two Gaussians, one to
the absorption and the other to the emission component, we measure equivalent
widths (EWs) of 0.42, 0.44, and 0.37~\AA\ for \caii\ $\lambda\lambda$8498,
8542, and 8662 (respectively).  In the absence of emission-line contamination,
the relative strengths of the three lines in old stellar populations are
approximately 0.4:1:0.9 (Terlevich, D\'\i az, \& Terlevich 1990).

Given these special circumstances, conventional methods of deriving stellar
velocity dispersions (e.g., Tonry \& Davis 1981) are of questionable utility in
this instance.  Instead, we adopt a more straightforward strategy, namely by
directly comparing the spectrum of NGC 4395 to diluted and velocity-broadened
spectra of template stars.  This method is similar to that discussed by Barth,
Ho, \& Sargent (2002), but here we limit the fitting only to the \caii\
$\lambda$8498 line, since it is least contaminated by emission (due to the
relatively large wavelength separation and the weakness of the adjacent Paschen
line; see Fig. 1).  Figure~2 compares the spectrum of NGC 4395 to those of four
K-giant stars\footnote{We did not observe any velocity template stars at Keck
in this setting.  These stars come from observations made with the Shane 3-m
telescope at Lick Observatory (Ho et al.  2003, in preparation); they have $R
\approx 50,000$.}  (luminosity classes I--III), whose \caii\ $\lambda$8498 line
has been diluted to match the strength in the galaxy.  Unlike weaker metal
lines, note that the intrinsic widths of the \caii\ triplets are {\it not}\
negligible at echelle resolution.  They range from $\sigma \approx 20-25$ \kms\
for K-type giants and bright giants to $\sigma \approx 30$ \kms\ for red
supergiants (Fig. 2; Ho et al.  2003, in preparation).  The dotted curve shows
that the diluted spectrum of HD 214868 (K2~III) convolved with a Gaussian with
$\sigma_*$ = 25 \kms\ provides a reasonable match to NGC 4395.  On the other
hand, we note that HD 206778 (K2~Ib), whose \caii\ $\lambda$8498 line has an
intrinsic $\sigma \approx 30$ \kms, also gives a decent fit, with {\it no}\
additional broadening.  Thus, we conservatively adopt $\sigma_* < 30 \pm 5$
\kms\ as the line-of-sight velocity dispersion of the nucleus, where the error
bar reflects our estimate of the probable systematic uncertainty.

\section{Size of the Nuclear Star Cluster}

  In order to determine the physical size of the region from which the stellar
Ca~II lines originate, we have used GALFIT (Peng et al. 2002) to decompose an
archival {\it Hubble Space Telescope (HST)}\ WFPC2 F814W ($I$-band) image
(GO-6464; obtained on 1998 January 6 UT) of the nuclear region of NGC 4395 into
several components, as shown in Figure~3. We find (1) an unresolved point
source with $I = 17.42 \pm 0.1$ mag, (2) a fairly round (axis ratio $0.83 \pm
0.01$, position angle $99^\circ$), compact source with a Sersic profile (index
= $2.63 \pm 0.05$), and (3) an extended component that can be fit with an
exponential profile.

   We identify the compact source, whose half-light radius is 0\farcs19 (3.9
pc), with a nuclear star cluster that produces the observed \caii\ lines; based
on the {\it HST} photometry, it contributes 58\% of the light in the effective
spectral aperture. The extended component accounts for only 4\% of the flux in
this aperture; accordingly, we neglect it here. Thus, we assign a radius of
$\sim 3.9$ pc to the stellar component detected in the Keck spectrum of 
the nucleus of NGC 4395. 

   The measured EW of the \caii\ $\lambda$8498 line, 0.42 \AA, is only about
25\% of that found in typical old stellar populations

\vskip 0.3cm

\psfig{file=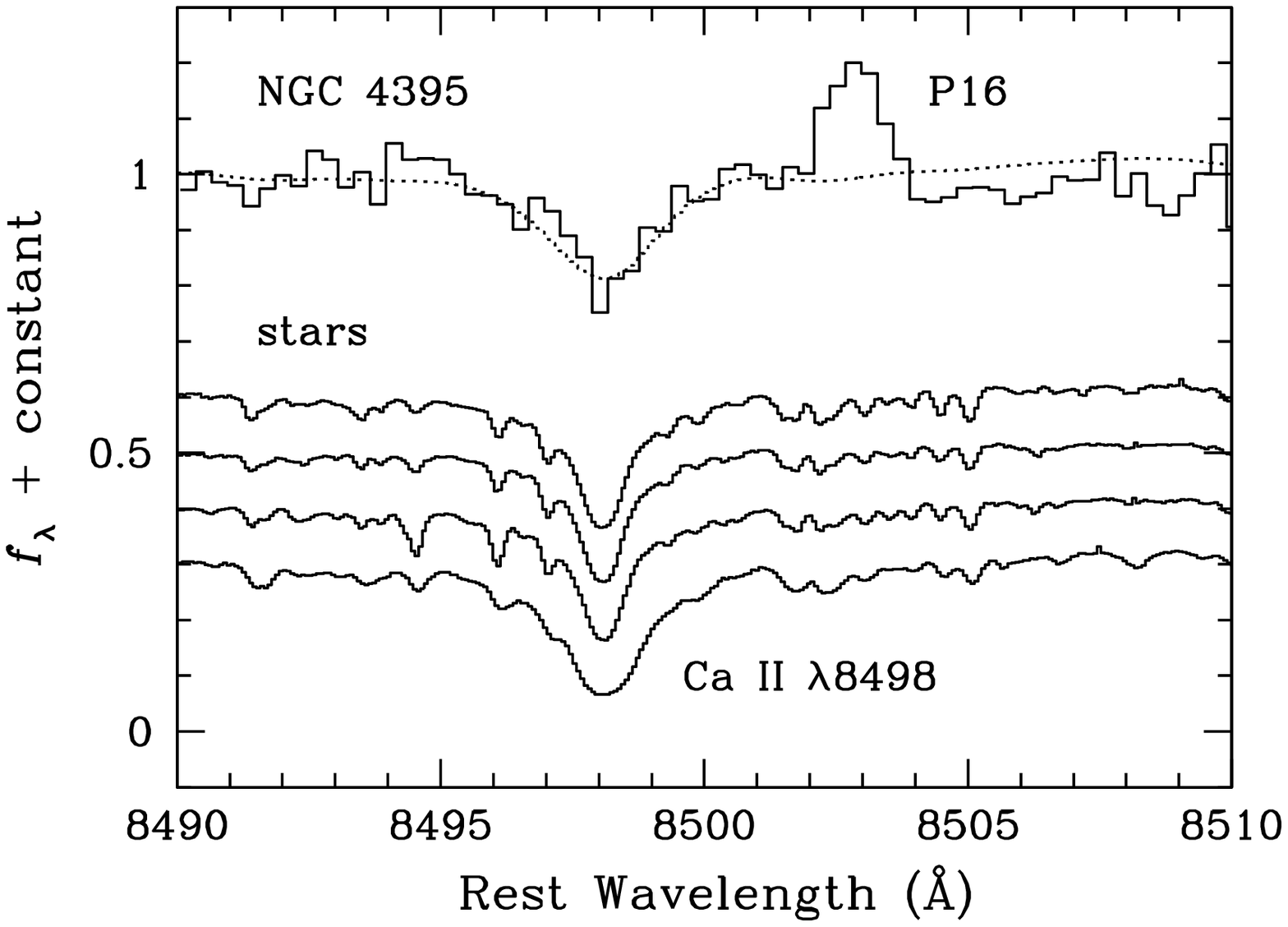,width=8.5cm,angle=0}
\figcaption[fig2.ps]{
Region around \caii\ $\lambda$8498 absorption and Paschen 16 emission for
the nucleus of NGC 4395 and four standard stars.  The continuum has been
normalized to unity.  From top to bottom, the stars are HD 213179 (K2~II),
HD 214868 (K2~III), HD 6953 (K7~III), and HD 206778 (K2~Ib).  The dotted curve
superposed on the spectrum of NGC 4395 is the spectrum of HD 214868 broadened
by $\sigma$ = 25 \kms.
\label{fig2}}
\vskip 0.3cm

\noindent
(Terlevich et al. 1990).
This may suggest that the nonstellar component contributes $\sim$75\% of the
$I$-band light, inconsistent with the smaller (38\%) contribution that we find
for the point source in the {\it HST} image. However, NGC 4395 has a very low
metallicity compared with typical spiral galaxies (e.g., Garnett 2002), and it
is well known that the \caii\ EW is small in stars having low metal abundance
($\sim$0.4--1.0~\AA\ for \caii\ $\lambda$8498; D\'\i az, Terlevich, \&
Terlevich 1989).  Roy et al. (1996) derive a value of log (O/H) = $-3.7$ for
an H~II region within $\sim 0.1$ kpc of the nucleus, and Kraemer et al. (1999)
suggest log (O/H) = $-3.5$ for clouds in the narrow-line region.  Since we do
not know the precise EW of the undiluted \caii\ $\lambda$8498 line in the
nuclear star cluster of NGC 4395, the amount of \caii\ dilution produced by the
nonstellar continuum is best estimated from the {\it HST} $I$-band image.

  Note that the low metallicity inferred for the central star cluster in NGC
4395 is not inconsistent with the high metallicities found in the broad-line
regions (BLRs) of high-redshift quasars (e.g., Pentericci et al. 2002, and
references therein). The stellar cluster in NGC 4395 is much larger (relative
to the probable black-hole mass) than the volume occupied by the BLR in
luminous quasars; moreover, there is a well-known quasar metallicity-luminosity
correlation, in light of which we expect low metal abundance even in the BLR of
NGC 4395.

\section{The Central Mass of NGC 4395}

The observed radial velocity dispersion of the stellar absorption lines can be 
used to set an upper limit to the virial mass of the central star cluster (or 
star cluster plus black hole).  For a bound system with negligible rotational 
support, $M \approx 2.5\langle \upsilon^2 \rangle r_h/G$ (Spitzer 1969), 
where $\langle \upsilon^2 \rangle$ is the mean-square velocity, $r_h$ is the 
half-mass radius, and $G$ is the gravitational constant.  Under the 
assumption of an isotropic velocity distribution, $\langle \upsilon^2 \rangle 
= 3 \sigma_*^2$, and so $M \approx 7.5\sigma_*^2 r_h/G$.  As discussed above, 
the most secure constraint on $\sigma_*$ is that of the \caii\ $\lambda$8498 
line: $\sigma_* < 30$ \kms. The best measurement on the size of the cluster
comes from the \hst\ image, which gives $r \approx 3.9$ pc (Fig. 3).  If we 
assume that light traces mass, then $r \approx r_h$, and 
$M$ \lax\ $6.2 \times 10^6$ \solmass.

\vskip 0.3cm

\psfig{file=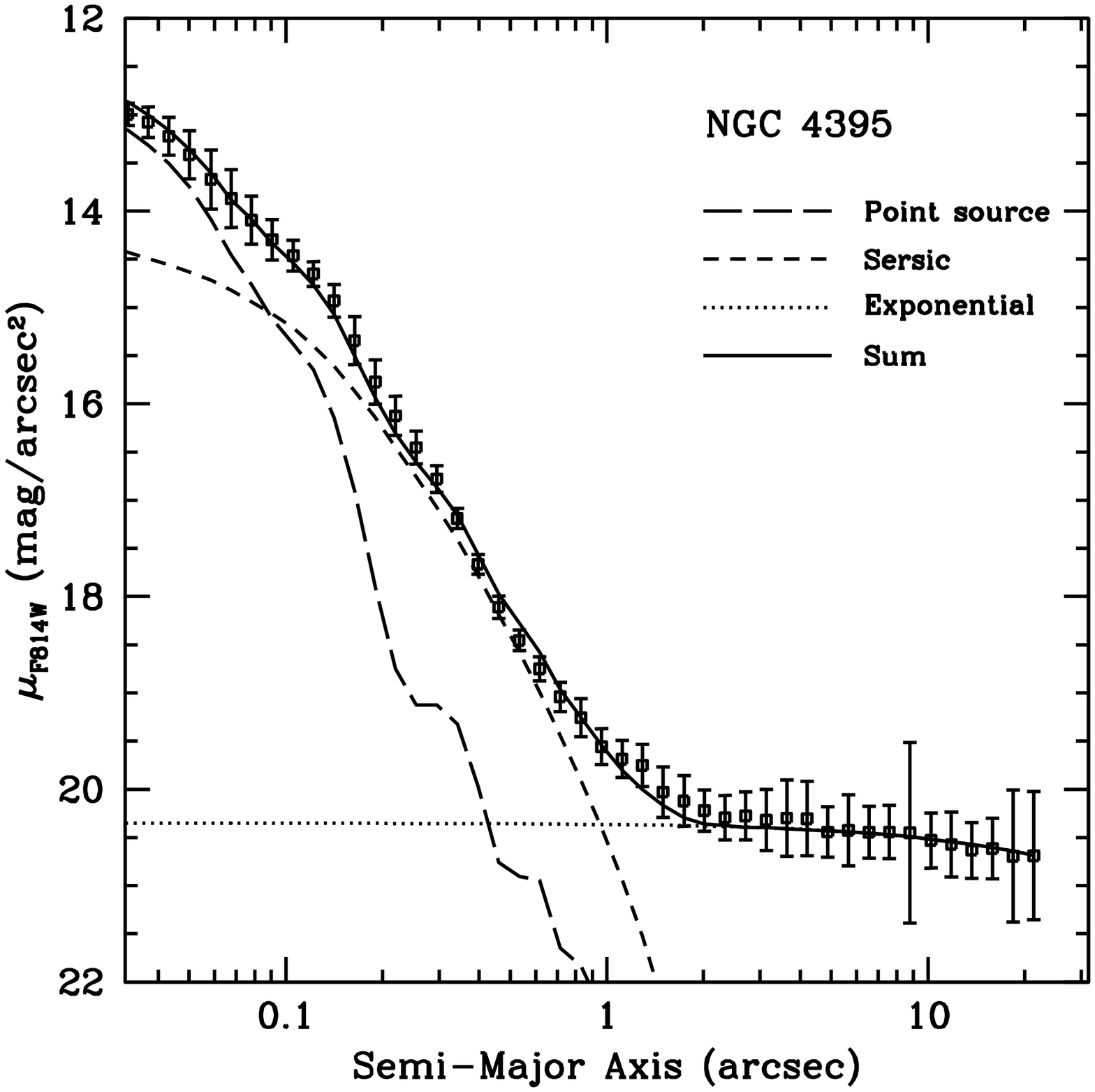,width=7.5cm,angle=0}
\figcaption[fig3.ps]{
Decomposition of the {\it HST}/WFPC2 F814W image of the nuclear region of NGC
4395. Data points are shown with error bars.  Undulations in the unresolved
point-source profile are an artifact of diffraction. The scale length of the
exponential disk may be erroneous by a factor of $\lesssim 2$; there was not
enough detector area to properly estimate the background sky.
\label{fig3}}
\vskip 0.3cm

The derived upper limit to the mass of the nucleus of NGC 4395 is a factor of
$\sim$10--100 larger than those of typical Galactic globular clusters (Mandushev,
Spassova, \& Staneva 1991), and is several times more massive than the largest
young ``super star clusters'' (SSCs) found in starburst systems (Ho \&
Filippenko 1996; Larsen et al.  2001; Smith \& Gallagher 2001). However, if the
actual mass is substantially less than the upper limit, and if only part of the
measured mass comes from stars (the rest being the black hole), then the true
mass may be comparable to that of the most massive globular clusters and
SSCs. Indeed, adopting an apparent $I$-band magnitude of 16.8 for the stellar
component of the nucleus (based on the photometric decomposition of the \hst\
image), $M_I \approx -11.3$ mag, within the realm 
of SSCs (e.g., Barth et al. 1995; Whitmore \& Schweizer 1995; Maoz et al. 2001)
and quite typical of nuclear star clusters in other late-type spiral galaxies
(B\"oker et al. 2002).  In the absence of reliable information on the age of
the stellar population in the nucleus of NGC 4395, it is difficult to judge
whether the inferred limit on the mass-to-light ratio [$M/L_I$ \lax\ $2
(M/L_I)_{\odot}$] is abnormally high or not.  The data, therefore, do not
require the presence of excess dark mass.  The nucleus of NGC 4395 may consist
of an essentially normal cluster of stars, with a small black hole at its
center accreting matter and producing the observed nonstellar activity. Indeed,
the AGN bolometric luminosity of $\sim 3 \times 10^{40}$ erg s$^{-1}$ (Moran et
al. 1999) can be produced by a $\sim 250~M_\odot$ black hole accreting at the
Eddington limit.

It is instructive to examine other indirect methods to estimate black hole
masses in AGNs. From photoionization modeling of the narrow-line region and
BLR in NGC 4395, Kraemer et al. (1999) find that the broad
H$\beta$ line originates from a region with a radius of $r_{\rm BLR}=3 \times
10^{-4}$ pc.  Assuming that the line-emitting gas is gravitationally bound,
that the orbits are randomly oriented, and that the observed width of the
H$\beta$ line traces the velocity dispersion of the gas, the virial mass
follows from $M_{\rm BH} \approx \upsilon^2 r_{\rm BLR}/G$.  Choosing $\upsilon
= (\sqrt{3}/2)$FWHM (Netzer 1990) and FWHM = 1500 \kms\ for H$\beta$ (Kraemer
et al. 1999), $M_{\rm BH} \approx 1.2 \times 10^5$ \solmass.

The size of the BLR can also be obtained through the luminosity-size relation 
recently established for type~1 AGNs studied with reverberation mapping.  
Equation (6) of Kaspi et al. (2000) relates $r_{\rm BLR}$ to the 
monochromatic luminosity of the featureless continuum at 5100 \AA, 
$\lambda L_{\lambda}(5100$~\AA).  From the spectral information given 
in Filippenko \& Sargent (1989) and Filippenko, Ho, \& Sargent (1993) 
[$F_{\nu} \propto \nu^{-1.6}$; $F_{\nu}(4400$~\AA) = 0.42 mJy], we find 
$\lambda L_{\lambda}(5100$~\AA) = $6.6 \times 10^{39}$ \lum\ and 
$r_{\rm BLR} = 3.3 \times 10^{-5}$ pc.  Although one might legitimately 
question whether the luminosity-size relation can be extrapolated to the 
ultra-low luminosity regime of NGC 4395, it is interesting to note that this
value of the BLR radius is a factor of 10 smaller than that obtained through 
photoionization modeling, remarkably consistent with a similar trend found in 
more luminous Seyfert~1 nuclei (Peterson 1993).  With the smaller value of 
$r_{\rm BLR}$, the virial mass is consequently an order of magnitude lower, 
$M_{\rm BH} \approx 1.3 \times 10^4$ \solmass.  

Finally, the black hole mass in NGC 4395 can be estimated from its X-ray 
variability characteristics.  Shih et al. (2003) applied the formalism of 
Hayashida et al. (1998) to {\it ASCA}\ observations of NGC 4395 and find 
$M_{\rm BH} \approx 10^4-10^5$ \solmass.  Although the systematic 
uncertainties of this method are not well understood, it is encouraging that 
the X-ray--based mass, within its relatively large allowable range, turns out 
to be consistent with the virial mass (calculated using $r_{\rm BLR}$ from 
the luminosity-size relation).   

To summarize, indirect estimates suggest that the putative black hole in NGC
4395 has a mass $M_{\rm BH} \approx 10^4-10^5$ \solmass.  If so, the AGN is
radiating at $\sim 2\times 10^{-2}$ to $2\times 10^{-3}$ of the Eddington
limit, consistent with the extrapolation of the mass-luminosity relation for
classical type 1 Seyfert nuclei and QSOs (Kaspi et al. 2000, eq. 11).  Thus,
the nucleus of NGC 4395 appears to be a low-mass analog of a QSO or type 1
Seyfert, with a ``normal'' accretion rate relative to the Eddington limit. By
contrast, LINERs have healthy-sized black holes ($M_{\rm BH}$ \gax\ $10^6$
\solmass) but very low accretion rates ($L_{\rm bol}/L_{\rm Edd} < 10^{-3}$; Ho
1999; Ho et al. 2000). This supports the hypothesis that the accretion rate is
one of the main factors in dictating the level of excitation exhibited by the
ultraviolet/optical spectrum: NGC 4395 has a high-excitation spectrum, as in
classical Seyferts and QSOs, while the spectra of LINERs are dominated by
low-excitation emission lines.

\section{Implications for Black Hole Demography}

The recent discovery of a tight correlation between black hole mass and the 
stellar velocity dispersion of the bulge (Gebhardt et al. 2000; Ferrarese \& 
Merritt 2000), the $M_{\rm BH}$--$\sigma_*$ relation, has inspired many 
efforts to couple the formation and growth of supermassive black holes to the 
formation of their host galaxies (e.g., Kauffmann \& Haehnelt 2000; Burkert 
\& Silk 2001).  An important clue is that the connection seems to involve 
primarily the bulge component, not the disk (Kormendy et al. 2003).  The most 
stringent limit to date comes from {\it HST}\ stellar-dynamical observations 
of the nucleus of the Scd galaxy M33, also a bulgeless system: Gebhardt et 
al. (2001) find that any central black hole would have to be less massive 
than 1500 \solmass.  The $M_{\rm BH}$ limit for M33 falls significantly 
below the mass predicted from the $M_{\rm BH}$--$\sigma_*$ relation.  

Do supermassive black holes {\it ever}\ form in pure disk systems?  Yes, as
illustrated by NGC 4395.  Our assessment in \S~5 suggests that the central
black hole in NGC 4395 is likely to have a mass between $\sim 10^4$ and $10^5$
\solmass, with a firm upper limit of $6.2 \times 10^6$ \solmass.  These mass
estimates are not inconsistent with the $M_{\rm BH}$--$\sigma_*$ relation of
Tremaine et al. (2002).  For $\sigma_* = 30$ \kms, it predicts $M_{\rm BH} =
6.6 \times 10^4$ \solmass.  Although current ground-based surveys indicate that
unambiguous type~1 Seyfert galaxies like NGC 4395 are quite rare, Ho et
al. (1997) show that more subtle signatures of nonstellar activity (based on
the spectrum of the narrow emission lines) can be found in $\sim$10\% of
late-type spirals (Sc and later).

\acknowledgments

The W.~M. Keck Observatory, made possible by the generous financial support of
the W.~M. Keck Foundation, is operated as a scientific partnership between the
California Institute of Technology, the University of California, and NASA. We
are grateful to the Keck staff for their assistance with the observations. We
thank Aaron Barth, Ryan Chornock, and Gibor Basri for useful comments, and
especially Chien Peng for producing Figure 3 and for doing the relevant
analysis. The excellent suggestions of an anonymous referee helped improve the
paper. Support for this work was provided by the NSF through grants AST-9417213
and AST-9987438, by NASA through grant NAG 5-3556, and again by NASA through
grant AR-7527 from the Space Telescope Science Institute, which is operated by
AURA, Inc., under NASA contract NAS5-26555.  The research of L.~C.~H. is
supported by the Carnegie Institution of Washington.


\end{document}